**Mapping nanoscale dynamic properties of suspended and supported multi-layer graphene membranes via contact resonance and ultrasonic scanning probe microscopies**

Marta Mucientes[1*], Robert McNair[2], Adrian Peasey[2], Shouqi Shao[2], Joshua Wengraf[2], Kunal Lulla[1], Benjamin J. Robinson[1,3] and Oleg Kolosov[1,3**].

[1]Physics Department, Lancaster University, Lancaster LA1 4YB, UK

[2]Graphene NowNANO CDT, School of Physics and Astronomy, The University of Manchester, Manchester, M13 9PL, UK

[3]Materials Science Institute, Lancaster University, Lancaster LA1 4YW, UK

Email: [*]m.sanjuanmucientes@lancaster.ac.uk, [**]o.kolosov@lancaster.ac.uk,

ABSTRACT. Graphene's (GR) remarkable mechanical and electrical properties - such as its Young's modulus, lowmass per unit area, natural atomic flatness and electrical conductance - would make it an ideal material for micro and nanoelectromechanical systems (MEMS and NEMS). However, the difficulty of attaching GR to supports coupled with naturally occurring internal defects in a few layer GR can significantly adversely affect the performance of such devices. Here, we have used a combined contact resonance atomic force microscopy (CR-AFM) and ultrasonic force microscopy (UFM) approach to characterise and map with nanoscale spatial resolution GR membrane properties inaccessible to most conventional scanning probe characterisation techniques. Using a multi-layer GR plate (membrane) suspended over a round hole we show that this combined approach allows access to the mechanical properties, internal structure and attachment geometry of the membrane providing information about both the supported and suspended regions of the system. We show that UFM allows the precise geometrical position of the supported membrane-substrate contact to be located and provides indication of the local variation of its quality in the contact areas. At the same time, we show that by mapping the position sensitive frequency and phase response of CR-AFM response, one can reliably quantify the membrane stiffness, and image the defects in the suspended area of the membrane. The phase and amplitude of experimental CR-AFM measurements show excellent agreement with an analytical model accounting for the resonance of the combined CR-AFM probe-membrane system. The combination of UFM and CR-AFM provide an beneficial combination for investigation of few-layer NEMS systems based on two dimensional materials.

### 1. Introduction

Graphene (GR) and other two-dimensional materials (2DM's) have drawn extensive interests thanks to their remarkable structures and exceptional mechanical, electrical, thermal and optical properties. GR in particular holds significant potential for micro and nanoelectromechanical systems (MEMS and NEMS) applications, transparent and flexible electrodes, optical modulators and nanomechanical resonators operating in classical and quantum regimes [1-5]. However, there are several major challenges for such devices; particularly important for NEMS applications, is the quality of the 2DM-substrate contact, effects of inhomogeneous in-plane stresses and, for the few-layer and heterostructured 2DM's, interlayer defects. To investigate and mitigate these, one needs matching nanoscale characterisation methods, with the ability to map nanomechanical properties, sensitivity to the stresses in the suspended areas, and the capability to evaluate the quality of the interfacial contact. While commercially available atomic force microscopy (AFM) have been commonly used to study suspended 2D materials [6-9], the material-substrate interaction and nanomechanical mapping across the whole NEMS structure remains challenging with limited reports in this area [10-12]. One of the promising techniques is to use contact resonance AFM (CR-AFM) shown to be able to detect subsurface holes under the 2DM's membrane[13-15] and to measure the local stiffness of the structure [15, 16]. At the same time, these studies indicated that the observed shape of the 2DM structure and recorded amplitudes varied significantly depending on the CR-AFM operation frequency, requiring independent methods to determine the geometry of such 2DM's "nano-drum". Here we combine AFM with ultrasonic excitation, ultrasonic force microscopy (UFM), and a phase sensitive variable frequency CR-AFM, to effectively observe the quality of the interface of the "supported" graphene and the substrate, to obtain nanomechanical maps of the suspended GR nanodrum structures and to provide quantitative comparison of the nanomechanical images and theoretical models.

**2. Experimental details**

The multi-layer graphene (MLG) nanodrums were produced via transfer of mechanically exfoliated MLG on top of a Si/SiO$_2$ substrate with 300 nm thermal oxide on the top. The substrate was pre-patterned with circular holes with a diameter of 1.9 µm etched to the depth of 150 nm in SiO2 layer *via* optical lithography and CHF$_3$/Ar reactive ion etching (see supplementary materials, SM, for the details). For the GR exfoliation [17], we used Gel-Pak® PF-4X film (0.5 mm thickness), with the resulting flakes transferred directly to the substrate. The substrate was treated in 98% Ar/2% O$_2$ plasma (PlasmaPrep2, Gala Instruments) for 10 min immediately prior to the transfer, to remove organic contamination and increase the adhesion of the graphene to the substrate.

The UFM and CR-AFM measurement setup was implemented by modifying a commercial AFM (MultiMode with Nanoscope-VIII controller, Bruker, Santa Barbara, USA). In both modes the out-of-plane vibration of the sample was realised by attaching it to a high frequency piezoceramic transducer (PI, Germany) *via* crystalline salol [18], with the transducer excited by a 33220A signal generator (Keysight, USA) (figure 1(a)). The resulting cantilever deflection signal was acquired using a custom-made signal access box with low noise high frequency signal buffers. In CR-AFM mode, the amplitude and phase of the AFM cantilever deflection at the excitation frequency in the range 10-100 kHz was detected by the lock-in amplifier (SRS-830, Stanford Research Systems, USA). In the UFM mode, the sample vibration was excited at the carrier frequency $f_{UFM}$ in the range 4 to 5 MHz and is amplitude modulated using a triangular waveform at frequencies $f_{mod}$ between 1-5 kHz. In this mode, the lock-in amplifier was detecting the amplitude of the deflection at the modulation frequency $f_{mod}$. As described elsewhere [18], the UFM signal at the modulation frequency is the result of nonlinear force-vs-distance dependence of the force interaction between the probe tip and the sample, reflecting the local stiffness of the material under the AFM tip, with UFM being especially sensitive to the materials with high surface stiffness such as graphene [19] and single or few layer 2DMs [20]. In both modes, a commercial contact AFM probe (ContAl-G, Budget Sensors) cantilever with the nominal 10 nm tip radius of curvature was used. The spring constant of the cantilever was measured using "Sader method" [21] to provide $k_c$ = 0.184 N m$^{-1}$, matching the data obtained by the thermal calibration method implemented by the AFM manufacturer (Bruker).

In CR-AFM mode, the cantilever is brought into the contact with the stiff material (SiO$_2$/Si substrate) resulting in the increase of the free cantilever resonance frequency of $f_0$=13 kHz to contact resonance frequency $f_{CR-R}$ = 64.6 kHz (in the contact with the substrate in the absence of lateral scanning). In order to find the resonance frequencies we performed frequency sweeps obtaining the maximum amplitude of the oscillatory deflection signal.

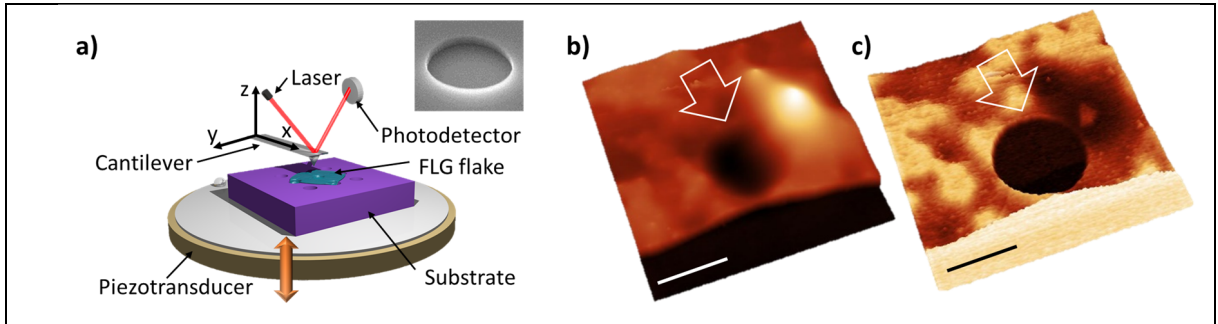

**Figure 1.** (a) Schematics of experimental setup for UFM and CR-AFM measurements, inset – SEM image of the hole in SiO2/Si substrate defined by optical lithography. (b) 3D rendering of contact AFM topography and (c) UFM nanomechanical image of the multilayer GR flake of 12 nm thickness over the hole in the substrate ($f_{UFM}$ =4.2 MHz). Scale bars 500 nm.

## 3. Results and discussion

The combination of AFM and high frequency excitation increases the sensitivity of the measurement setup to the mechanical properties compared to the standard force-distance spectroscopy and its option Peak Force™. These allowed mapping of the entire sample area, with true nanoscale spatial resolution and, in case of UFM, eliminating friction[22] providing a non-destructive approach especially valuable when the supported sample is investigated. Figure 1 b, c) shows a typical topography and UFM images of the GR flake spanning the hole in the Si/SiO2 substrate (inset in figure 1a), with the UFM image showing a well-defined hole in the supported area. The figure 2 compares with the sample while scanning across the sample, to quantify the sample contact stiffness, $k_s$ using analytical approaches described elsewhere[23, 24]. Figure 2(a) shows the topography of the test sample with the 12 nm MLG flake deposited over the hole. In the topography image, the hole is barely observed as the smooth depression on the surface of MLG (arrow). The UFM nanomechanical image figure 2(b) clearly delineates the boundary of the hole corresponding to the edge of the suspended region enabling high precision measurements of the geometry of the membrane boundary (radius $R$=940± 5 nm in this structure). It also reveals neighbouring defects corresponding to the weaker MLG – substrate interaction and interfacial defects (darker contrast in figure 2(b)). While UFM has superior contrast to the stiff (supported) areas of GR nanostructures, it is not sensitive to the variations of the contact stiffness within the suspended membrane area, which in this case is below the UFM sensitivity range of $10^2$ to $10^5$ N m$^{-1}$ [25, 26]. It should be noted, though, that for 2D membranes of smaller dimensions and/or larger thickness, the UFM may become directly sensitive to the properties of the suspended areas as well. In the case of our nanostructure, to map and quantify its stiffness, we used CR-AFM that provided an excellent mechanical contrast in the suspended areas. Figure 2 panels (c-h) shows a set of the CR-AFM amplitude and phase images obtained by detection of amplitude and phase of the cantilever deflection at the different sample excitation frequencies $f_{CR}$[27]. For reference, the topography and UFM image of the same area are provided in the panels (a) and (b). While CR-AFM provides excellent contrast to the suspended area of the membrane and to the supported areas of the MLG, one can see that both the contrast, shape, and dimensions of the membrane are varied significantly depending on the frequency used.

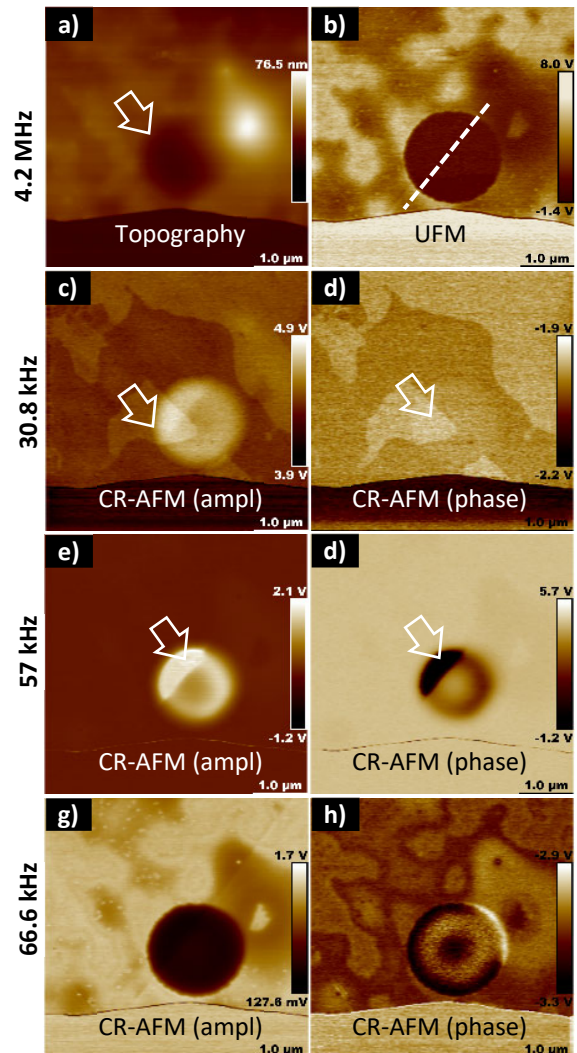

**Figure 2.** (a) AFM topography and (b) UFM map of the 12 nm thick MLG flake on the substrate with the hole with $R$=940 nm. (c-h) Set of the CR-AFM amplitude and phase maps of the same area (c-d) at 30.8 kHz, (e-f) 57.0 kHz, and (g-h) 66.6 kHz. Arrows highlight particular features detailed in the text.

In figure 2 we compare three sets of the CR-AFM phase and amplitude response for the excitation frequency around the tip-substrate contact resonance frequency ($f_{CR-R}$=64.6 kHz). The $f_{dr}$=30.8 kHz amplitude image (figure 2(c)) shows the hole as a donut-shaped bright area with a darker central area, whereas the phase image (figure 2(d)) has practically no contrast to the hole. At the same time, both amplitude and phase images clearly show the internal structure of the membrane not visualised in the topography, UFM or CR-AFM at other frequencies. Given the shape of the features and the absence of the contrast in the topographical image (sensitive to the surface features) and in UFM (that would be most sensitive to sample – interface contact [10]) these are interpreted as an internal crack in the MLG flake (arrows in figure 2(c,d)) which induces a change in the stiffness of this particular area. At increased frequency $f_{CR}$=57.0 kHz (figure 2(e,f)) both amplitude and phase CR-AFM clearly shows the similar donut-shaped hole, now with the reduced diameter, with a clear segment (bright in the amplitude and dark in the phase image) running at 45° across the hole, another internal feature of the membrane. Finally, a CR-AFM image at $f_{CR}$=66.6 kHz (figure 2(g,h)) shows the amplitude image as an uniformly dark area, similar to UFM image in size, whereas the phase is presented as a concentric bright halo around the darker hole. The amplitude images are generally similar to ones reported by Ma et al [13] whereas phase images require more detailed analysis. This frequency also shows MLG-substrate contact variations similar to the UFM image, although with opposite contrast. While the rich contrast obtained in CR-AFM for such GR nanostructure is outstanding, and ability to observe subsurface features corresponds to the one reported by the variable frequency AFM imaging of cellulose fibres [28], the strong qualitative and quantitative dependence of the contrast on the excitation frequency required more detailed experimental studies and matching analysis to help with the interpretation of the CR-AFM data.

In order to consistently analyse frequency dependent response of CR-AFM, we obtained a set of one dimensional scans – profiles along the single line shown in figure 2(b), while changing the frequency in 1 kHz steps from 50 to 69 kHz that includes the reference, $f_{CR}$, (stiff sample). By continuing acquisition of the amplitude and phase signal, we collected the two-dimensional amplitude and phase – vs position/ frequency graphs, correspondingly, $A(r, f_{dr})$ and $A(\phi, f_{dr})$, as shown in figure 3 where the horizontal axis is the position of the probe across the hole, $r$, and vertical – the driving frequency $f_{dr}$, whereas the brightness corresponds to the amplitude or phase of the CR-AFM.

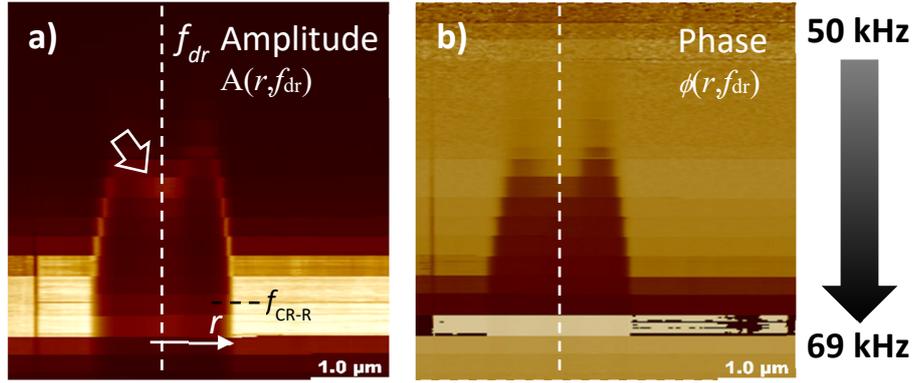

**Figure 3.** Plot of the CR-AFM response as the function of the position *r* across the hole (horizontal axis, *r*) and frequency (vertical axis, $f_{dr}$) with (a) amplitude A as the function of A(*r*,$f_{dr}$) and (b) $\phi$(*r*,$f_{dr}$) phase profile at driving frequency $f_{dr}$ changing from 50 to 69 kHz in 1 kHz steps.

As we can see, the apparent "diameter" of the hole as seen by the CR-AFM is varied from the largest at $f_{dr} = f_{CR-R}$ to significantly reduced at lower driving frequencies ($f_{dr} < f_{CR-R}$). Also, the second maximum (arrow in the figure 3(a)) that would create a "ring" in the x-y image, appeared in the amplitude CR-AFM at the frequencies $f_{dr} < f_{CR-R}$. These phenomena can be qualitatively explained by the changing of contact stiffness of the tip-membrane contact that is the lowest in the centre and the largest at the periphery approaching "infinite" stiffness for the supported MLG layer. It is known that the resonance frequency of the cantilever-tip in contact with the membrane decreases with the decrease of its stiffness [23, 24]. Therefore the condition for the resonance for the decreased driving frequency $f_{CR}$ are satisfied in the more central area of the membrane away from its border, resulting in the smaller diameter of the membrane, creating a bright ring at the position where the resonance occurs. In order to better understand and quantify our observations, we have used the analytical and modelling approach described below.

First, we estimate the resonance frequencies of mechanical resonators based on suspended MLG flakes. These depend on GR Young's Modulus, $E_s$; Poisson's ratio, $v_s$; density, $\rho_s$; pre-tension, *T*, as well as the flake thickness, *t*, and the radius of the suspended area, *R*. For the resonators made by graphene monolayer the tension usually dominates as their bending rigidity is insignificant comparing with one produced by the tension [6]. However, for the resonator based on thicker MLG flake, the minimal resonance frequency at zero tension is determined by the circular plate behaviour, with tension further increasing this frequency. For the plate the fundamental resonance frequency is given by the equation [29]

$$f_p = \frac{10.21}{4\pi}\sqrt{\frac{E_s}{3\rho_s(1-\nu_s^2)}}\frac{t}{R^2} \qquad (1)$$

For MLG flake with the dimensions presented in the figure 2, the fundamental resonance frequency $f_p$~138 MHz. This is several orders or magnitude higher than the frequencies of both CR-AFM (10-100 kHz) and UFM (4-5 MHz), used in our experiments, and therefore we can considered the GR membrane as the spring, neglecting its mass, allowing to use the formalism reported elsewhere [23, 30, 31] to describe the dynamics of the cantilever terminated with the spring. The motion of the cantilever-sample system can be modelled as a cantilever supported at the free end by a spring, with respective $k_c$ and $k_s$ cantilever and sample stiffness, $k_s$ = d$F$/d$z$ where $F$ is the force experienced by the cantilever and $z$ is the deflection of the drum from the equilibrium measured at the point of contact.

The equation of the dynamic deflection of the cantilever is given as $\frac{\partial^2 z}{\partial t^2} + \frac{E_c h^2}{12\rho_c}\frac{\partial^4 z}{\partial x^4} = 0$ where $E_c$ is the Young's Modulus, $\rho_c$ is the density, $h$ is the thickness of the cantilever and $x$-$z$ are the coordinates according the sketch in the figure 1(a). With density and Young's modulus of Si known, the length of the cantilever can be easily determined, by the optical microscope image, however the thickness of the cantilever cannot be determined with the reasonable precision. We therefore used the "Sader" model developed elsewhere [23] to find and experimentally measured cantilever stiffness $k_c$ = 0.184±0.008 N m$^{-1}$ that allows us to estimate the thickness of the cantilever $h$ = 5.86 10$^{-7}$ m. We can then calculate the resonance frequency of the cantilever in the contact with the samples of varied sample stiffness $k_s$ (see SM for the derivation of this dependence). By substituting the values for the free cantilever, $k_s \rightarrow 0$, and experimentally obtained resonance frequency: $f_0$=13.8 kHz and the experimentally obtained CR-AFM frequency for the Si/SiO2 substrate that can be presented as infinitely stiff for the fundamental mode of the contact resonance [23], $k_s \rightarrow \infty$, in the substrate area of $f_{CR-R}$ = 64.6kHz, we obtain the CR-AFM response curve in figure 3. This curve 3 relates the resonant frequency of the cantilever with the ratio between the stiffness of the sample and the cantilever ($k_c/k_s$).

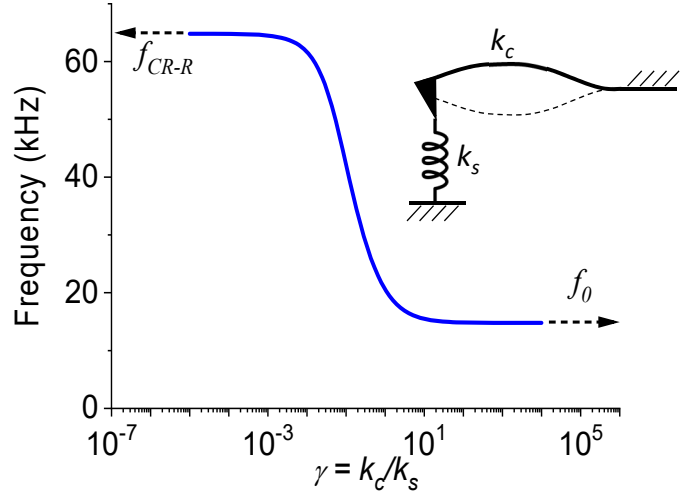

**Figure 4.** Resonant frequency for the CR-AFM $f_{CR}$ for the first contact resonance mode (n=1) as a function of the cantilever-sample stiffness ration $k_c/k_s$.

In the case of a circular plate, the stiffness of the sample probed by the AFM tip depends on the distance from the centre of the plate to the contact point, as well as on the boundary conditions at the edge of the MLG flake. The two typical boundary conditions are – a) simply supported at the single circular line around the hole rim resulting in no bending moment applied to the plate at the edges (figure 5(a)) and b) - clamped when the edges of the plate are forced to be kept parallel to edges of the substrate surface (figure 5(b)). In the first case, the distribution of the stiffness as a function of radius $r$ from the centre of the plate is given by [32]

$$k_s(r) = \frac{1}{H(r)c_0 a^3} \tag{2}$$

where $a$ is the radius of the plate and $H(r)$, $b_0$ and $c_0$ expressed as

$$H(r) = \frac{1}{\pi D a^4}\frac{2(1+\nu_s)}{9(5+\nu_s)}(r^3 - b_0 a r^2 + c_0 a^3); \quad b_0 = \frac{3(2+\nu_s)}{2(1+\nu_s)}; \quad c_0 = \frac{4+\nu_s}{2(1+\nu_s)} \tag{3}$$

with the bending stiffness $D$ defined as $D = \frac{E_s t^3}{12(1-\nu_s^3)}$.

In case of MLG flake clamped in the edges (figure 5(b)), the reference [32] provides stiffness distribution governed by the following equation

$$k_s(r) = \frac{16\pi D a^2}{(a^2 - r^2)^2} \tag{4}$$

In the centre of the plate the stiffness is then expressed as $k_s(r=0) = 16\pi D/a^2$. By substituting the parameters of the flake into the equation and using the in-plane Young's modulus of graphene of 1 TPa [33], we obtain the calculated value of the stiffness in the centre of the plate of $k_s$ = 5.94 N m$^{-1}$, that is more than twice the experimental one, $k_s$ = 2.15±0.008 N m$^{-1}$. At the same time, for the simply supported plate, the stiffness in the centre will be expressed as

$$k_s(r=0) = \frac{9\pi D(5+\nu_s)}{2c_0^2 a^2 (1+\nu_s)} \quad (5)$$

resulting in calculated $k_s(r=0)$=2.35 N m$^{-1}$. We can therefore conclude that the simply supported model has better fit with the experimental results.

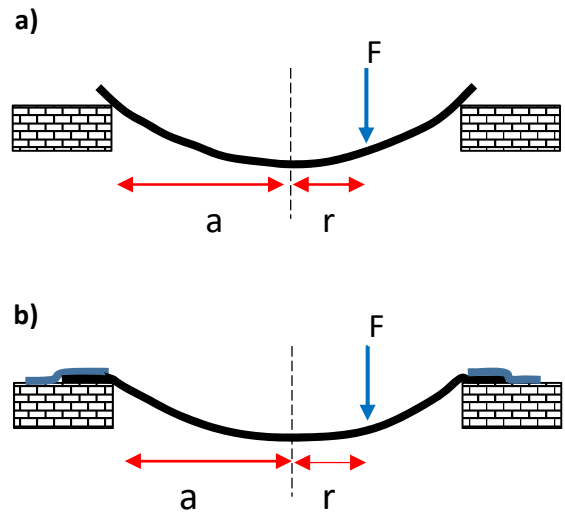

**Figure 5.** Two typical boundary conditions for the plate suspended over circular hole – (a) simply supported edges and (b) non-clamped edges.

The equations 2 and 3 together with the relation presented in the figure 4 (with the corresponding equations in SI) allow to link the resonant frequency $f_{CR}$ and the stiffness of the membrane with the spatial position (the radius from the centre where the probe is located) where the maximum amplitude is located. We can further extend this analysis by considering the amplitude and phase of a cantilever as driven damped simple harmonic oscillator with the resonance frequency calculated by the approach defined above. In this case, the amplitude $A$ and phase $\varphi$ of the response of the cantilever are expressed as

$$A = \frac{A_0 [\omega_{CR}(r)]^2}{\sqrt{([\omega_{CR}(r)]^2 - \omega_{dr}^2)^2 + \frac{\omega_{dr}^2}{[\omega_{CR}(r)]^2 Q}}}; \quad \varphi = \tan^{-1}\frac{\omega_{CR}(r)\omega_{dr}}{Q([\omega_{CR}(r)]^2 - \omega_{dr}^2)} \quad (6)$$

where $A_0$ is the amplitude in the resonance, $\omega_{CR}(r)=2\pi f_{CR}(r)$ is the circular resonance frequency of the cantilever in the contact that depends on the position $r$, and $\omega_{dr}=2\pi f_{dr}$ is the circular driving frequency, and $Q$ is the quality factor of the cantilever in the contact. Using this formalism, we can now simulate the amplitude dependence of the CR-AFM $A(r,f_{dr})$ that was presented in figure 3(a) and compare it to the experimental data. The results of this simulation are presented in the figure 6(a,b) and show an excellent agreement between the model and the experimental measurements. As one can see, at $f_{dr}$ close to the frequency of contact resonance of the supported MLG $f_{CR-R}$ (65 kHz, figure 6(d)) all suspended areas have low response, whereas at the frequency below $f_{CR-R}$ (62 kHz, figure 6(c)) the resonance appears at the radius close to the rim – explaining the bright ring we observed in the figure 2(g). Essentially, by comparing the resonance frequency of the supported and the suspended areas figure 6(e), the modelling presented in the figure 4, and calibrated cantilever stiffness, allows to directly determine the MLG plate stiffness at 2.15±0.008 N m$^{-1}$. By using real-time mapping of the resonance frequency at each point of the sample during scan (similar to approach used in the piezo-force microscopy [34]) ad analysis above, it is possible to produce high precision stiffness maps allowing direct interpretation of the internal structure in the multilayered and heterostructured 2D materials observed in the figure 6(c-h).

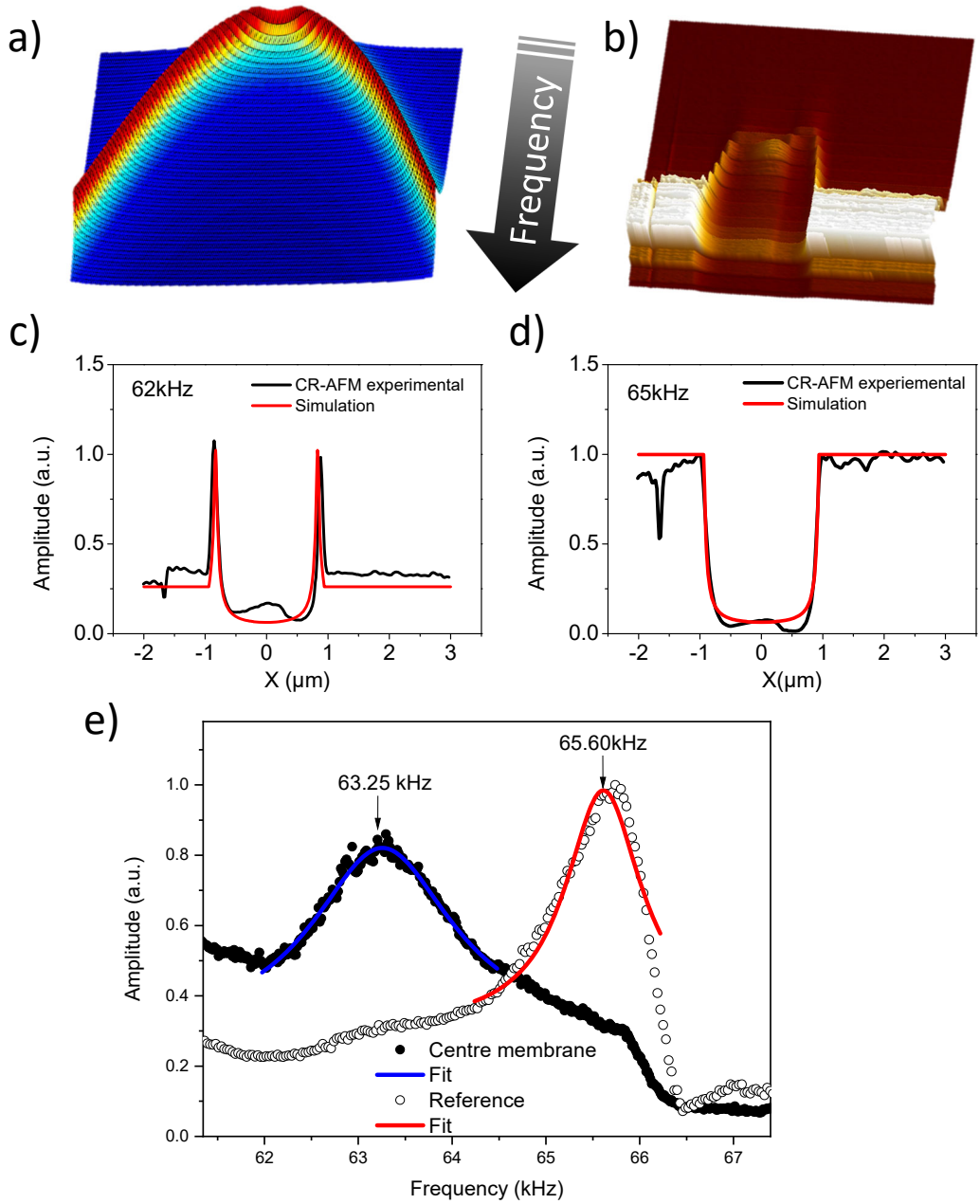

**Figure 6** (a) Modeling of the amplitude response A(*r*,*f*$_{dr}$) of the CR-AFM as the function of the radius from the centre of the plate and the driving frequency *f*$_{dr}$ for the 50 to 69 kHz frequency range, (b) experimental CR-AFM data of the A(*r*,*f*$_{dr}$) response for the same frequency range. (c, d) comparison of the simulated and measured one-dimensional profiles *A(r)* at the frequencies below and above contact resonance frequency for the solid contact. (e) Experimentally measured CR-AFM response in the centre of the MLG plate vs supported graphene, the frequency shift allows precisely determine the stiffness of the GR membrane.

## 4. Conclusions

In summary, here we use the combination of the low and high frequency excitation dynamic force microscopy in the frequency range from 10 kHz to 5 MHz to investigate the details of the nanomechanical behaviour of the multilayer graphene plate suspended on the micromachined substrate. In particular, the high frequency UFM mapping allowed precise determination of the geometry of the suspended region even for the relatively thick plate, as well as observation of the faults at the MLG-substrate interface. By building an analytical model of the CR-AFM in application to the elastically deformed circular plate, we were able to perform absolute measurements of the mechanical stiffness of the 2D material nanostructure and to interpret the novel contrast phenomena specific to the 2D materials. The excellent agreement between the experimental and modelling data and high sensitivity of the method to the internal structure of MLG opens a great possibility to investigate multilayer heterostructures of 2D materials, and the 2D material –substrate interface.

**Acknowledgements**

Authors acknowledge the support of Manchester-Lancaster EPSRC Graphene NowNANO CDT, Lancaster University QTC and Lancaster Materials Science Institute. OVK acknowledges support of EPSRC grants EP/K023373/1 and EP/G06556X/1, EU grant QUANTIHEAT and Paul Instrument Fund, c/o The Royal Society.

# Supplementary Materials.

**Mapping nanoscale dynamic properties of suspended and supported multi-layer graphene membranes via contact resonance and ultrasonic scanning probe microscopies**

Marta Mucientes[1*], Robert McNair[2], Adrian Peasey[2], Shouqi Shao[2], Joshua Wengraf[2], Kunal Lulla[1], Benjamin J. Robinson[1,3] and Oleg Kolosov[1,3**].

[1]Physics Department, Lancaster University, Lancaster LA1 4YB, UK

[2]Graphene NowNANO CDT, School of Physics and Astronomy, The University of Manchester, Manchester, M13 9PL, UK

[3]Materials Science Institute, Lancaster University, Lancaster LA1 4YW, UK

Email: [*]m.sanjuanmucientes@lancaster.ac.uk, [**]o.kolosov@lancaster.ac.uk,

## 1. Substrate preparation.

The substrates were first cleaned by sequential sonication in acetone and isopropyl alcohol (10 min each) followed by an $O_2$ (40 sccm; 200 mTorr, 50W) plasma in the reactive ion etching (RIE) (Oxford Instruments PlasmaPro® NGP80) during 3 minutes to remove any organic contamination from the surface. To pattern the substrates, the negative photo-resist S1813 was spin-coated at 4000 rpm for 45 sec and baked at 115° for 2 min. The resist was exposed to the UV light during 2.7 seconds with hard contact in the MJB4 Mask Aligner (SUSS MicroTec, Germany) to transfer the pattern. Post-exposure, the samples were baked for 1 min at 115° and developed in MFCD26 for 90 seconds. The next step was 7.5 min RIE, with a gas mixture of $CHF_3$ (25sccm) and Ar (5 sccm), at a pressure of 30mTorr in the chamber and a RF power of 150 W. Finally, the wafers were cleaned to remove all the resist residue with the $O_2$ plasma during 3 minutes and in the ultrasonic bath with Acetone and IPA during 5 minutes. Figure 1 shows the resulting hole.

## 2. Resonance frequencies of clamped plate and membrane.

To find the fundamental resonance frequency $f_{plate}$ of a clamped plate of thickness *t* and radius *R*, we used the equation

$$f_{plate} = \frac{10.21}{4\pi}\sqrt{\frac{E_s}{3\rho_s(1-\nu_s^2)}}\frac{t}{R^2} \qquad (1)$$

Nevertheless, for the determination of the fundamental resonance of a membrane $f_{membrane}$ under pre-tension $T$, the equation used was:

$$f_{membrane} = \frac{2.4048}{2\pi R}\sqrt{\frac{T}{\rho_s t}} \tag{2}$$

In both cases, the physical properties of the material are density $\rho_s$, Young's Modulus $E_s$, and the Poisson's ratio $v_s$.

### 3. Calculating the resonance of the cantilever in contact with the spring.

The motion of the cantilever-sample can be modeled as a cantilever supported in the free end by a spring, with respective $k_c$ and $k_s$ cantilever and sample stiffness. The equation of the deflection of a cantilever of thickness $h$ is given as:

$$\frac{\partial^2 z}{\partial t^2} + \frac{E_c h^2}{12 \rho_c}\frac{\partial^4 z}{\partial x^4} = 0 \tag{3}$$

where $x$-$z$ are the coordinates, $E_s$ and $\rho_s$ are respectively the Young's modulus and density of the cantilever material.

In the literature, they assumed that the solution of the equation (3) takes the form

$$z = C\sin(\omega t + \delta)\Phi(x) \tag{4}$$

with

$$\Phi(x) = (\sin\alpha + \sinh\alpha)\left(\cos\frac{\alpha}{L}x - \cosh\frac{\alpha}{L}x\right) \\ + (\cos\alpha + \cosh\alpha)\left(\sin\frac{\alpha}{L}x - \sinh\frac{\alpha}{L}x\right) \tag{5}$$

where $\omega$ is the angular frequency, $L$ is the length of the cantilever and the constant

$$\alpha = \left(\frac{12\rho_c \omega^2 L^4}{E_c h^2}\right)^{\frac{1}{4}} \tag{6}$$

relates the cantilever parameters with the resonance frequency. This constant was obtained from the boundary conditions. Furthermore, considering that the sample behaves such as linear spring, the motion equation can be written as

$$\frac{k_c}{3k_s}\alpha^3 P(\alpha) = Q(\alpha) \tag{7}$$

where $P(\alpha)$ and $Q(\alpha)$ are defined as

$$P(\alpha) = 1 + \cos\alpha\cosh\alpha \tag{8}$$

$$Q(\alpha) = \cos\alpha \sinh\alpha - \sin\alpha \cosh\alpha \qquad (9)$$

Solving numerically the equations 8 and 9, for the boundary conditions below of the relation between the stiffness of the sample and the cantilever, we calculated the value of $\alpha$ = 3.92.

$$k_s \ll k_c \Rightarrow P(\alpha) \to 0 \qquad (10)$$
$$k_c \ll k_s \Rightarrow Q(\alpha) \to 0 \qquad (11)$$

In the first case, the ratio between the stiffness of the sample and the cantilever corresponds experimentally with free cantilever, $k_s \to 0$, with resonance frequency: $f_{free\text{-}cantilever}$ = 13.8 kHz. And the second one is corresponding with the probing point in the supported MLG $k_s \to \infty$, $f_{substrate}$ = 64.6 kHz.

Solving numerically the equations 8 and 9, for the boundary conditions below of the relation between the stiffness of the sample and the cantilever, we calculated the value of $\alpha$ = 3.92.